%% file: main.tex
\documentclass[a4paper, amsfonts, amssymb, amsmath, reprint, showkeys, nofootinbib, twoside,aps,prl]{revtex4-2}
\usepackage[english]{babel}
\usepackage[utf8]{inputenc}
\usepackage[colorinlistoftodos, color=green!40, prependcaption]{todonotes}
\usepackage{amsthm}
\usepackage{mathtools}
\usepackage{physics}
\usepackage{xcolor}
\usepackage{colortbl}
\usepackage{graphicx}
\usepackage[left=23mm,right=13mm,top=35mm,columnsep=15pt]{geometry} 
\usepackage{adjustbox}
\usepackage{placeins}
\usepackage[T1]{fontenc}
\usepackage{lipsum}
\usepackage{csquotes}
\usepackage{todonotes}
\usepackage{natbib}

\usepackage{balance}
\begin{document}
\title{Ultrafast electrical control of dipolariton-based optical circuits with a few femto-joul per bit power consumption}

\author{%
    Dror Liran,$^{1}$ 
    Kirk Baldwin,$^{2}$
    Loren Pfeiffer,$^{2}$
    Hui Deng$^{3}$
    and Ronen Rapaport$^{1}$
}
\email{ronenr@phys.huji.ac.il}
\affiliation{$^{1}$Racah Institute of Physics, The Hebrew University of Jerusalem, Jerusalem 9190401, Israel}
\affiliation{$^{2}$Department of Electrical Engineering, Princeton University, Princeton, NJ, 08544, USA}
\affiliation{$^{3}$Department of Physics and Department of Electrical Engineering and Computer Science, University of Michigan, Ann Arbor, MI 48109, USA}
\date{} 

\begin{abstract}
The next generation of photonic circuits will require programmable, ultrafast, and energy-efficient components on a scalable platform for quantum and neuromorphic computing. Here, we present ultrafast electrical control of highly nonlinear light-matter hybrid quasi-particles, called waveguide exciton-dipolaritons, with extremely low power consumption. Our device performs as an optical transistor with a GHz-rate electrical modulation at a record-low total energy consumption $\sim$3 fJ/bit and a compact active area of down to 25 $\mu$m$^2$. This work establishes waveguide-dipolariton platforms for scalable, electrically reconfigurable, ultra-low power photonic circuits for both classical and quantum computing and communication.
\end{abstract}

\maketitle
\section{Introduction} 
Integrated photonic circuits are the most promising platform for the rapidly developing fields of optical information processing \cite{Rudolph2017WhyComputing, Bogaerts2020ProgrammableCircuits, McMahon2023TheComputing, Elshaari2020HybridCircuits,Dietrich2016GaAsCircuits}, including classical and quantum simulators and computers \cite{Berloff2017RealizingPolaritonsimulators,Alyatkin2024AntiferromagneticLight}, as well as neuromorphic calculators \cite{Shastri2021PhotonicsComputing,Ballarini2020}. A crucial requirement of an integrated photonic platform \cite{Shastri2021PhotonicsComputing, Bartolucci2023Fusion-basedComputation} is the ability of ultrafast \textit{electrical} control of individual nodes with ultra-low power consumption per node per operation.

Exciton polaritons (EP) \cite{Deng2010}, resulting from the strong coupling of confined photons and two-dimensional excitons, are excellent building blocks for photonic circuits. They can be directly addressed due to their photonic part, and they have stronger optical nonlinearities due to their excitonic part.
Recent demonstrations of ultrafast all-optical control of EP in microcavities, including optical switching \cite{Ballarini2013, SpinSwitchAmo2010, Sanvitto2016,Marsault2015,Sturm2014,Ballarini2013,Gao2012PolaritonSwitch,SpinSwitchAmo2010,Sturm2014,Sannikov2024RoomGates,Masharin2024GiantMetasurfaces}, EP condensate-based photonic lattice simulators \cite{Berloff2017RealizingPolaritonsimulators,Alyatkin2024AntiferromagneticLight,Jacqmin2014DirectPolaritons}, and EP based neuromprphic computing \cite{Ballarini2020,Opala2024RoomCrystal} are just a few examples of the promise of EP platforms.
\begin{figure*}
    \centering
    \includegraphics[width=0.9\linewidth]{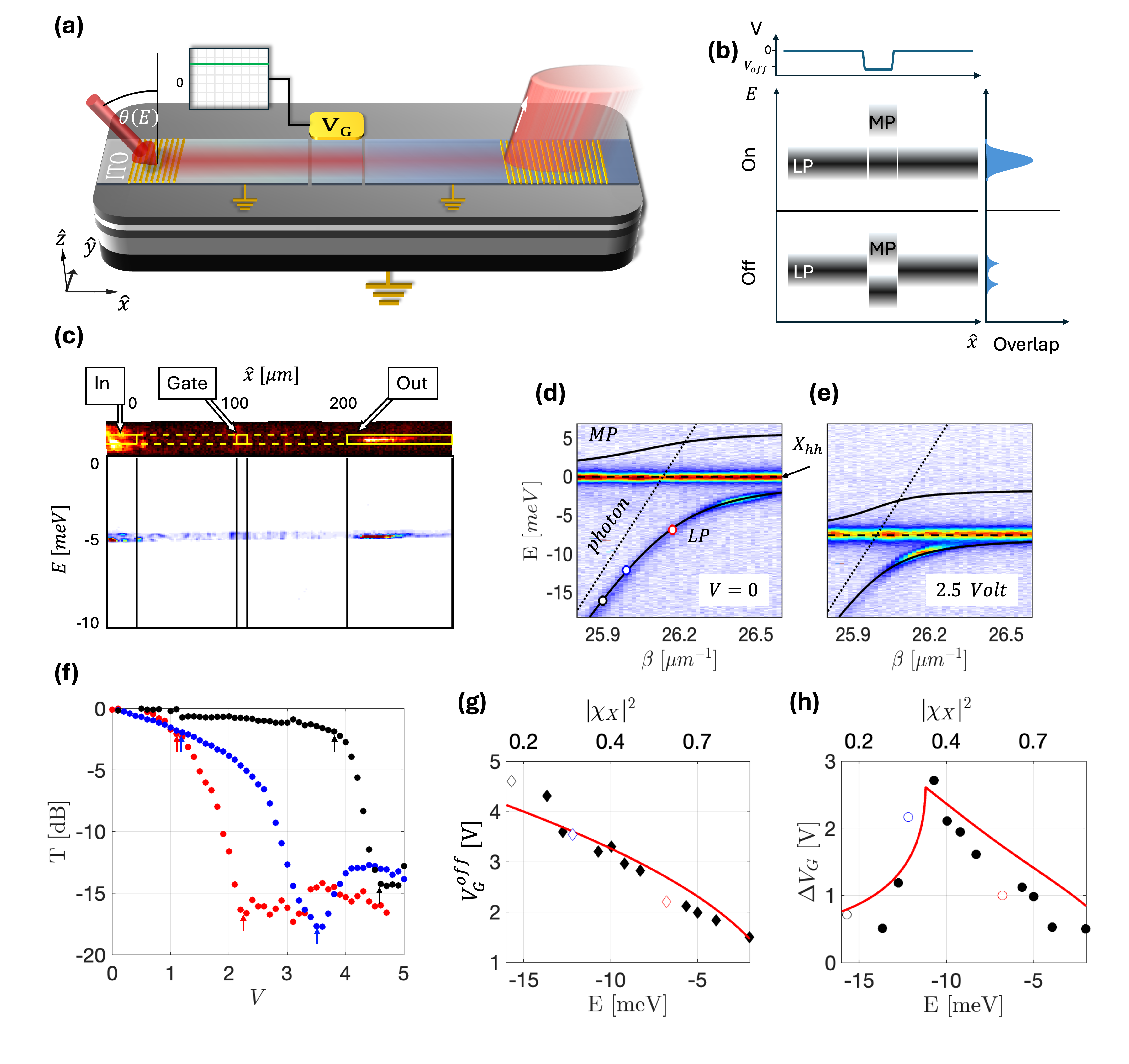}
    \caption{\textbf{Setup and electric field dependence of experiments with gated-DWEP devices: }
    \textbf{(a)} Schematic diagram of the device structure, with the ITO electrode divided into three parts: the gate \( V_G \) and the unbiased channel. The incident laser excites the sample at \( \theta(E) \) to satisfy the WEP dispersion. The transmitted light is a function of the field amplitude \( V_G \).
    \textbf{(b)}
    Energy diagram for the polariton switching. Top Panel: schematic of the voltage values along the waveguide. Middle panel: "on"  - the gate is flat with the channel allowing the polaritons to transmit. Bottom panel: "off" - the gate is biased such that the states from the channel meet the LP-MP gap in the gate, and the polaritons do not transmit.
    \textbf{(c)}
    Top panel: a real-space image of the WEP experiment. The laser is injected in the left grating, and the WEP signal couples out from the right grating.
    Bottom panel: a spectrally resolved imaging measurement of the WEP.
    \textbf{(d) and (e)} measured WEP Dispersion at \( V_G = 0 \) and \( V_G = 2.5 \, V/\mu m \), respectively. The plots show the energy \( E \) as a function of the wavevector \( (\beta) \). The lower polariton (LP) and middle polariton (MP) model fits are indicated by solid black lines. A flat dashed line marks the heavy-hole exciton, and a dotted diagonal line marks the bare WG-photon. The colored dots in (d) correspond to the three curves in (f).
    \textbf{(f)} The transmission in the channel as a function of \( V_G \) for three different WEP excitation energies \(E = -7,-12,-16 meV\) marked in (d) respectively. The arrows mark the points defined as the on-off transition.
    \textbf{(g)} The voltage value (\(V_G\)) of the first minimum in the transmission. The values for the curves in (f) appear in the corresponding color. The red line plots the voltage value required for the exciton to shift such energy.
    \textbf{(h)} The voltage difference (\(\Delta V_G\)) between the first minimum in transmission and a transmission of \(0.6\). The two colored points represent the corresponding transitions marked in figure (f) by colored arrows. The red lines mark the model prediction.
    }
    \label{fig:Fig1}
\end{figure*}

Yet, for modern complex photonic circuitry, highly desirable are electrically controlled photonic nodes that are densely integrated into monolithic chip geometries. For microcavity-based EP elements, only optical control has been demonstrated so far; 
significant optical power is required per node, typically exceeding $10^3$J/node/operation at 1GHz \cite{Ballarini2013}; scaling up layers of nodes for deep circuits is also a major challenge, as microcavities cannot be easily stacked.

Alternatively, monolithic chip-integrated waveguide exciton-polaritons (WEP) \cite{Walker2013,Rosenberg2016} are well suited for scalable EP circuits. Several key elements have been demonstrated recently \cite{Liran2018a,Walker2017,Suarez-Forero2021,Rosenberg2018a,LiranPRX,Nigro2022IntegratedInterferometry}.
An important step toward locally reconfigurable and scalable waveguide exciton-polariton (WEP) circuits is the realization of \textit{electrically gated dipolar WEP} (DWEP) structures, where a top gate applies a perpendicular electric field to wide quantum wells (QWs) inside the waveguide, inducing a quantum-confined Stark effect and voltage-controlled exciton dipoles \cite{Rosenberg2016}. These dipoles exhibit strong dipole–dipole interactions and screening effects, enhanced further by the ultralight effective mass of the polaritons \cite{Christensen2024MicroscopicInteractions}, resulting in record-high effective nonlinearities—up to two orders of magnitude larger than in unpolarized polaritons \cite{Rosenberg2018a}—and enabling key demonstrations such as electrically controlled few-photon transistors \cite{LiranPRX}. Remarkably, electrically tunable quantum correlations and a partial 2-photon blockade were recently demonstrated, showing potential for realizing DWEP for electrically switchable, universal 2-photon gates \cite{Ordan2024ElectricallyRadii}.

Here, we demonstrate a fast, sub-nanosecond temporal control of such nonlinear DWEP-switches operating at record-low powers of sub 3 Femto-Joule/bit and a very small footprint of $< 50 \mu m^2$ per node, a significant step towards complex optical circuitry based on WEP, towards universal photonic computation.

\begin{figure}
    \centering \includegraphics[width=0.9\linewidth]{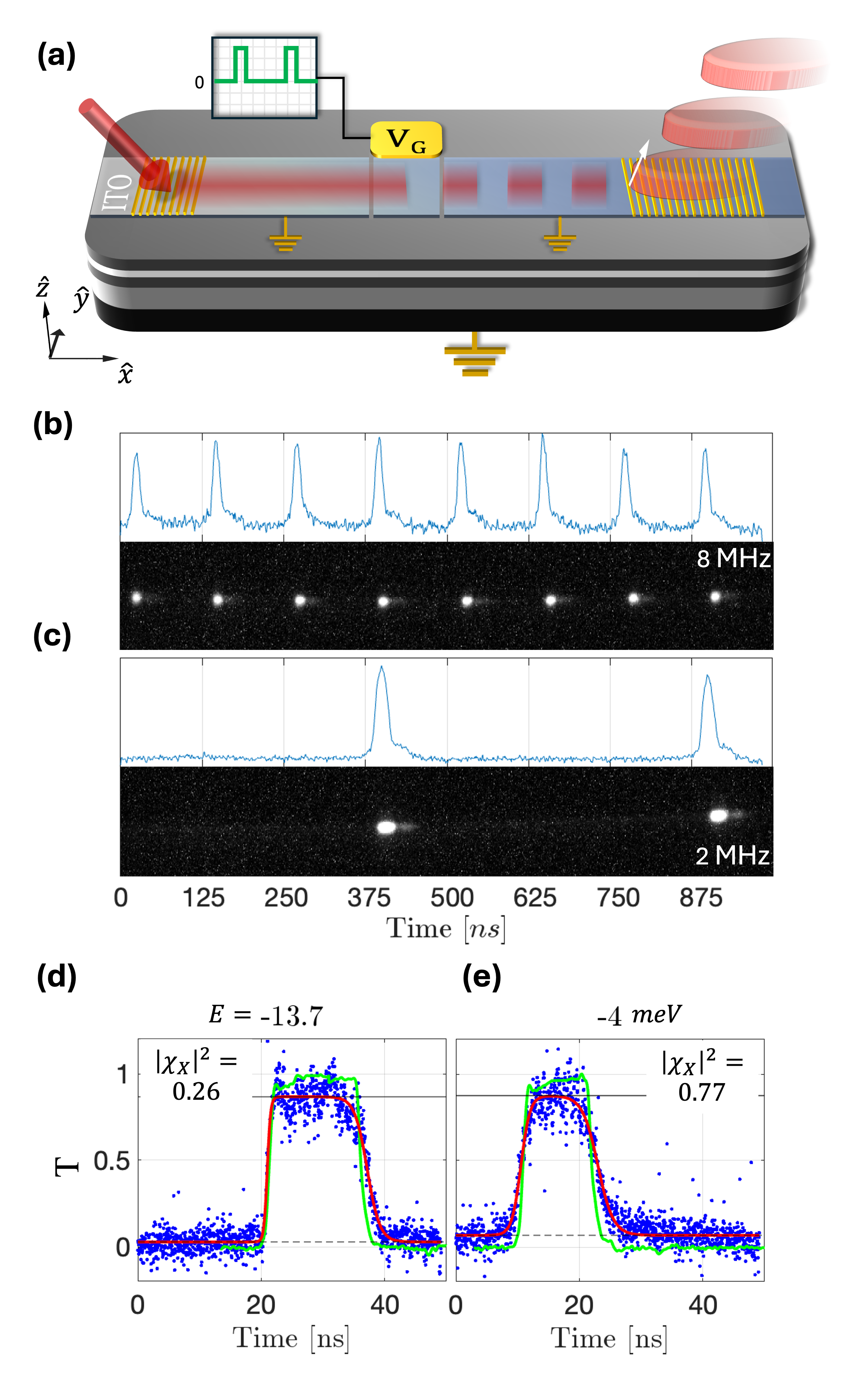}
    \caption{\textbf{Pulse Modulation: }
\textbf{(a)} Illustration of the device response to a square voltage modulation of the gate.
\textbf{(b) and (c)} Time-resolved measurement of the transmitted signal at 8 and 2 MHz, respectively. The bottom panel plots the raw data from the streak camera, while the top panel plots the integrated amplitude.
\textbf{(d) and (e)} Normalized time-domain pulse transmission (blue dots) for two different polariton energies \( E = -13.7 \, \text{meV} \) and \( E = -4 \, \text{meV} \), plotted along the normalized electric signal input. The data is fitted to a pulse function (red line). The black lines mark the "on" (solid) and "off" (dashed) states.
}
    \label{fig:Fig3}
\end{figure}

\section{DWEP device characterization}
The device used in our experiments (Fig. \ref{fig:Fig1}(a)) is a 200 \(\mu m\) long and $5$ \(\mu m\) wide waveguide channel, optically defined by an indium tin oxide (ITO) strip, with two Au diffraction gratings for input and output at either end. The waveguide itself is constructed out of an $Al_{0.4}Ga_{0.6}As$ core with 12 embedded 20nm-wide $GaAs$ quantum wells (QWs). For details on the full sample fabrication, see Ref. \cite{LiranPRX}.
The strong interaction of the transverse-electric (TE) polarized waveguide mode with the heavy-hole ($hh$) and light-hole ($lh$) excitons leads to the formation of three polariton modes: Lower-Polariton (LP), Middle-Polariton (MP), and Upper-polariton (UP) \cite{Rosenberg2016}.  Each polariton mode is a superposition of the bare TE-photon and the two excitons, represented as $\ket{\psi_{pol}(\beta)}^i = \chi_{ph}^i(\beta) \ket{\psi_{ph}} + \chi_{hh}^i(\beta) \ket{\psi_{hh}} +  \chi_{lh}^i(\beta) \ket{\psi_{lh}}$, where $i=$ LP, MP, and UP respectively, and the Hopfield coefficients satisfy \(|\chi_{hh}|^2+|\chi_{lh}|^2+|\chi_{ph}|^2=1\). The total exciton fraction of the polariton is defined here as $\abs{\chi_{X}(\beta)}^2 = 1-\abs{\chi_{ph}(\beta)}^2 $. LP-polaritons having $\ket{\psi_{pol}(\beta)}^{LP}$, $E_{LP}(\beta)$ can be excited at the left grating by a resonant laser (Ti:Sa, CW) with its energy and incidence angle matching the desired position on the polariton dispersion. We define a relative DWEP dispersion $E(\beta)=E_{LP}(\beta)-E_{hh}$, measured with respect to the unbiased $hh$-exciton energy, $E_{hh}$. Such dispersions are shown in Fig. \ref{fig:Fig1}(a, d). The output signal is collected from the right grating and imaged onto a spectrometer. 
Cross-polarization is employed in both the excitation and emission paths to isolate the polariton emission from the scattered laser light. Additionally, spatial filtering further reduces laser scattering (see Fig. \ref{sm:fig:setup} in the SM).
To allow independent control of the electric field in each section, the ITO strip that defines the optical waveguide \cite{LiranPRX} is divided into three sections by a $<1 \mu m$ gap in the electrode. The middle section, named the "gate", is electrically biased. The gate is \(10 \mu m \) long and is centered between the input and output gratings. The outer sections (hereafter the "channel") are held at zero bias.

When a field is applied to the gate, the DWEP transmission between the input and output gratings decreases due to the reduced overlap of the DWEP density of states under the gate and in the channel \cite{LiranPRX}; an illustration of the mechanism is presented in Fig. \ref{fig:Fig1}(b), where minimum transmission occurs when the gate voltage $V_G$ is shifting the LP energy by exactly one Rabi frequency $\Omega(V_G)$, i.e., $\Delta E(\Delta V_G) = \Omega (V_G)$, where (\(\Delta V_G\)) is the voltage difference for switching between on and off states. Fig. \ref{fig:Fig1}(f) plots the DWEP transmission $T$ (normalized to its maximum) as a function of the gate voltage (\(V_G\)) for three different excitation energies \(E = -7,-12,-16 meV\). Arrows on each curve indicate the \(T=0.6\) (-2.2 dB) and minimal transmission points.

Such transmission curves are then used to extract $V_G(E)$ required for blocking the DWEP propagation for different excitation energies, as plotted in Fig.\ref{fig:Fig1}(g).From the same transmission curves we also extract \(\Delta V_G(E)\), Fig. \ref{fig:Fig1}(h). Interestingly, \(\Delta V_G(E)\) shows a non-monotonic behavior.
This behavior arises from the nonlinear dispersion of the DWEP, $E(\beta)$. For simplicity, we employ a two-mode WEP model, which allows us to derive the following expression:
\begin{equation}
    \Delta V_{G}= V_G^{off}\left (1-\sqrt{\frac{-(E(\beta)_{0} +\Omega(V_G))}{-E(\beta)_{0}}} \right)
    \label{Eq:DF}
\end{equation}
Where $E(\beta)_{0}\equiv E(\beta,V_G=0)$ is the polariton energy which the laser resonantly injects into the sample, $V_G^{off} = \sqrt{\frac{-E(\beta)_{0}}{\alpha'}}$ is the voltage required to bring the exciton to the injection energy, and $\alpha' = 1.53 meV/V^2$\cite{LiranPRX} is the electric polarizability of the QW hh-exciton.
The model, presented in detail in the SM, agrees well with the data, as is shown by the red lines in Fig.\ref{fig:Fig1}(g,h).\\

\begin{figure}[]
    \centering
    \includegraphics[width=0.9\linewidth]{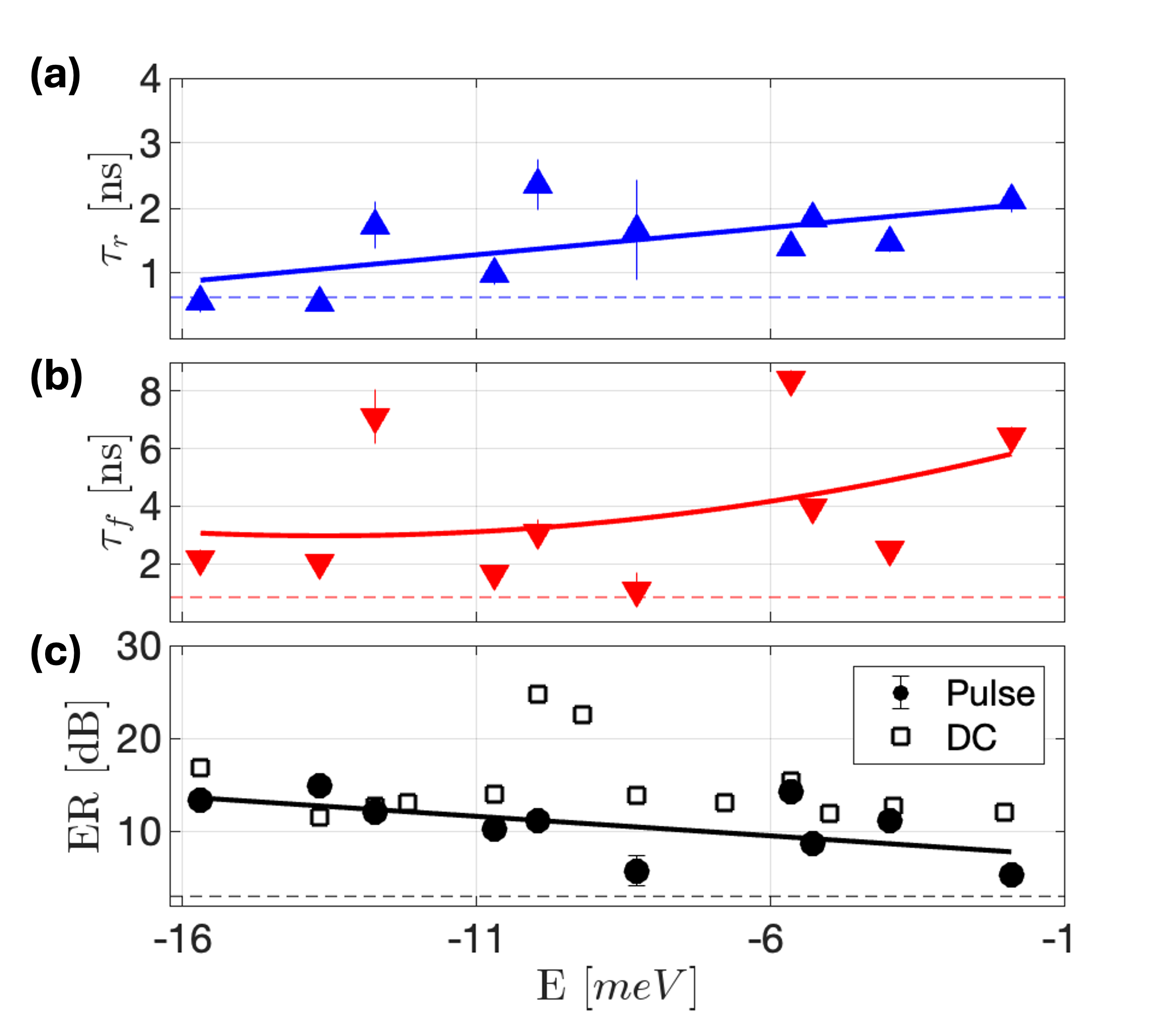}
    \caption{\textbf{Bandwidth, Extinction, and Power consumption: }
    \textbf{(a) and (b)} Extracted transition times (rise and fall) as a function of the excited WEP energy. The lines are guides to the eye.
    The error bars represent the fit uncertainty values.
    The dashed line marks the input voltage transition time.
    \textbf{(c)} Extinction ratios as a function of excited WEP energy. The line is a guide to the eye. A dashed black line marks the minimal required ER (3 dB).
    The corresponding DC extinction ratio is also plotted with empty squares.
    }
    \label{fig:Fig4}
\end{figure}

\setlength{\tabcolsep}{0pt} 

\begin{table*}[ht]
    \centering
    \textbf{Table I: Integrated electro-optic modulators}
\begin{tabular}{ccccc}
    \hline 
         Mechanism \hspace{0.5cm} &Material \hspace{0.5cm} & Power  consumption \hspace{0.5cm} & ER \hspace{0.5cm} & Footprint\\
         & &[\(fJ/bit\)] @ 1 GHz$^*$ & [dB] &[\(\mu m ^2\)]\\
         \hline
         \rowcolor{gray!20} DWEP (this work) &GaAs& \(\sim 3\)$^{**}$ & 25$^{***}$  &  25$^{\dagger}$\\
         Pockels effect (MZI$^{\dagger\dagger}$) \cite{Wang2018IntegratedVoltages} &LN & 26 &  30 & \(10^4\)\\
         Carrier injection \cite{Reed2010SiliconModulators}$^{\ddagger}$ &Si & 500  & 8 &   \(10^3\)\\
         Plasmonic–organic–hybrid (MZI) \cite{Haffner2016PlasmonicMicroscale} & Si and Au & 2750 
         &  25 \footnote{loss of 5dB per element} & \(\sim 10\)\\
         MC EP (all optical) \cite{Ballarini2013} & GaAs$^{\ddagger\ddagger}$ & 1980
         & 12 &  $1.6\times10^3$ \\
                 
         \hline 
    \end{tabular}
    \begin{flushleft}
    \small{$^*$ The power consumption was rescaled to a bit rate of 1GHz with duty-cycle of 0.5}
    \\
    \small{$^{**}$ see the SM for a full derivation}
    \\
    \small{$^{***}$ See DC measurements of ER in Fig. \ref{fig:Fig4}, limited by the SNR}
    \\
    \small{$^{\dagger}$ projected}
    \\
    \small{$^{\dagger\dagger}$ Mach-Zender Interformeter}
    \\
    \small{$^{\ddagger}$ The device with the lowest power consumption}
    \\
    \small{$^{\ddagger\ddagger}$ Can also be realized in many other material systems, see Ref. \cite{Sanvitto2016}}
    \end{flushleft}
\end{table*}

\section{Electrical modulation of DWEP}
to test the frequency response of the system, we injected WEP resonantly through the input grating using a CW laser, while modulating the gate voltage using a square electrical pulse $f(t)$ with a nanosecond rise and fall times. The experiment was repeated with different periods and duty cycles, as depicted schematically in Fig. \ref{fig:Fig3}(a). The field amplitudes and offsets of the electrical modulation were selected based on the DC measurements presented in Fig. \ref{fig:Fig1} (f-h), and can be represented by \(V(t) = V_{G}^{off}+\Delta V_G \times f(t)\). The output signal was imaged onto a streak camera. Such Streak images for two different modulation frequencies are presented in Fig. \ref{fig:Fig3}(b,c).
Fig. \ref{fig:Fig3}(d,e) presents the time-domain DWEP transmission of single pulses (blue dots) plotted on top of the modulating electric field (solid green line) for WEPs excited at \(-13.7,-4 meV\), corresponding to excitonic fractions of \(|\chi_X|^2 = 0.26,0.77\) respectively. The black lines indicate the “on” (solid) and “off” (dashed) states. The transmission data is fitted with a pulse function (red line):
\begin{equation}
    T(t) = T_{on}/2\times\left(\tanh(\frac{t-t_r}{\tau_r})-\tanh(\frac{t-t_f}{\tau_f})\right)+T_{off}
    \label{eq:pulse}
\end{equation}
where \(\tau_r\), \(\tau_f\) are the rise and fall times, respectively. 

\paragraph{Bandwidth:} From such fitting as above, we extracted \(\tau_r\), \(\tau_f\), as well as the extinction ratio (ER), $ER=T_{on}/T_{off}$, for various excitation energies, for electrical pulses with a 4MHz carrier frequency, and pulse durations of $15-20$ ns. The extracted rise and fall times are plotted in Fig. \ref{fig:Fig4}(a,b), and are both smaller for lower energy DWEPs, corresponding to more photon-like polaritons. Rise times as short as $0.5$ns and fall times as short as $1$ ns are measured. As seen from Fig. \ref{fig:Fig3}(d,e), the optical signal of photon-like DWEPs follows the electrical pulse. This means that the electrical pulse generator still limits the rise and fall times and not intrinsically by the WEP system. Therefore, we conclude that our devices can operate at $>$GHz bandwidth.

\paragraph{Extinction ratio:} the extinction ratio is shown for various DWEP energies in Fig. \ref{fig:Fig4}(c,d).
Again, more photon-like DWEPs tend to have a higher extinction ratio, indicating more efficient switching. Importantly, these photon-dominated states also have faster propagation in the WG, lower propagation loss, and better in- and outcoupling efficiencies. $ER$ values as high as 15dB are measured, limited only by the SNR of the Streak camera. To demonstrate that the actual ER values are higher, we plot the $ER$ for the DC case, which had a better SNR of up to $25$ dB.

\paragraph{Energy consumption:} the energy consumption for our device is composed of two factors, the energy of photons in a cycle and the electrical energy to operate the switching, given by the current-voltage product in a cycle.
To estimate this, we measured the current of the device during operation, with $V_G^{off} = 3.1 V$ and $\Delta V_G$ of 2.25, we measured currents of $I = $ 0.46 and 0.85 $\mu$A, respectively. For a duty cycle of $R=50\%$ and frequency of $f=1$ GHz, the energy consumption per operation is $1.3 fJ/bit$; the measurement details appear in the SM.
Optically, the optical power required for the nonlinear DWEP-transistor operation is 1.2 fJ per cycle at 1 GHz is required; (see SM).
Finally, electrical and optical energy consumption adds up to about $2.5 fJ$ per cycle at 1 GHz for a fully functional electrically-modulated transistor device.\\

\section{Discussion} 
We demonstrated DWEP signal modulation with a bandwidth exceeding 1 GHz, primarily constrained by the limitations of the electrical pulse generator and the electronic circuitry, together with a very high extinction ratio. A proper electrical design can significantly improve the bandwidth, as was demonstrated in  GaAs-based electro-optical devices that have achieved modulation frequencies up to \(100 GHz\) \cite{GaAsHighSpeed1994}.
With a minimal footprint (down to 25 \(\mu m^2\)), such elements are well suited for high-density photonic circuitry. The footprint could be further reduced by shortening the gate length to a few times the WEP wavelength in the medium (\(\lambda \simeq 240 nm\)), and decreasing the channel width to \(\sim 0.5 \mu m\) by side etching \cite{Liran2018a,Suarez-Forero2021}, resulting in an overall footprint as small as \(\sim 1\mu m^2\) per node.

Remarkably, our device demonstrates overall power consumption smaller than \(3 fJ/bit\), which, as far as we know, sets a record compared to other platforms, as detailed in the table.
We emphasize that low power consumption per node per operation is a crucial factor in large-scale fast circuitry, including those designed for classical neuromorphic computing as well as for quantum computing. Both cases require a large amount of reconfigurable elements \cite{deCea2021EnergyConsumption}. A further reduction in power consumption can be achieved by minimizing tunneling between the top and bottom contacts, as well as reducing the surface area of the electrical contacts, which in our current design constitute the most significant contributors to leakage currents.

Finally, the same device displays an electrically tunable, optical nonlinearity which is, as far as we know, the highest of any exciton-polariton-based system, and is comparable to that of Rydberg atoms \cite{Ordan2024ElectricallyRadii}. As was mentioned before, this led to a demonstration of a fully operational optical transistor operating with 1.2 fJ/bit \cite{LiranPRX}, and to quantum correlations at the two-photon level \cite{Ordan2024ElectricallyRadii}, which can be controlled very accurately by the applied electric field. The current demonstration of local >GHz modulation of an optical node, will allow a reconfigurable polariton-based optical circuitry, where the linear and nonlinear function at each node can be controlled separately and be reconfigured for each operation, making electrically polarized DWEP an excellent candidates for reconfigurable deep optical circuits for either neuromorphic or quantum processors in a monolithic platform.
Low propagation loss is also essential for a deep photonic circuit, particularly at the quantum limit where signal amplification is impossible. Our current device has a propagation loss of approximately 4-10 dB/cm\cite{Liran2018a}, with a lower loss for high photonic fractions, which is a typical value for photonic losses in AlGaAs WG systems. The photonic loss can be significantly reduced to 0.4 dB/cm by integrating AlGaAs on an insulator (AlGaAsOI) \cite{Chang2020Ultra-efficientMicroresonators}. Given the achievable element density of AlGaAsOI, the effective loss per number of elements compares with state-of-the-art silicon-on-insulator (SOI) systems.
Finally, for full integration, the chip should also incorporate single-photon detectors, which have already been realized in similar AlGaAs systems using superconductor-nanowire single-photon detectors (SNSPDs) \cite{Dietrich2016GaAsCircuits,Digeronimo2016IntegrationGaAs}.

\paragraph{Funding.}
R.R. and D.L. acknowledge the support from the Israeli Science Foundation Grant 1087/22, and from the NSF-BSF Grant 2019737. H.D. acknowledges the support of the National Science Foundation under grant DMR 2004287, the
Army Research Office under grant W911NF2510055, the Air Force Office of Scientific Research under grant FA2386-21-1-4066, and the Gordon and Betty Moore Foundation under
grant N031710. This research is funded in part by the Gordon and Betty Moore Foundation’s
EPiQS Initiative, Grant GBMF9615 to L. N. Pfeiffer, and by the National Science Foundation
MRSEC grant DMR 2011750 to Princeton University

\vfill
\bibliography{references}
\bibliographystyle{unsrt}
\balance
\newpage
\input{SM.tex}
\end{document}

%% file: SM.tex
\clearpage
\renewcommand{\thefigure}{S\arabic{figure}}
\renewcommand{\theequation}{S\arabic{equation}} 
\renewcommand{\thetable}{S\arabic{table}} 
\setcounter{figure}{0}
\setcounter{page}{1}
\setcounter{table}{0}
\setcounter{section}{0}

{\newpage \huge Supplementary Information}
\\
\\
The Supplementary Information contains supporting calculations, a description of the optical setup, and raw data Images for DC and pulsed experiments.

\section{Voltage modulation for DWEP Switching - model} \label{sm:sec:DF}
Here we develop a simple model for the voltage modulation required for switching between full transmission $T_{on}$  minimal transmission $T_{off}$. Minimal transmission occurs at a gate voltage amplitude $V_G$ that shifts the energy of the incoming WEP with a given $\beta$, $ E_{LP0}(\beta)$ under the gate into the middle of the LP-MP gap \cite{LiranPRX}, which corresponds to:
\begin{equation}
    E_{LP0}\simeq E_{hh} - \alpha' (V_G^{off})^2,
\end{equation} 
and
\begin{equation}
    E_{LP0}+\Omega(V_G^{on})\simeq E_{hh} - \alpha' (V_G^{on})^2,
\end{equation} 
where $\alpha' = 1.3/0.85 = 1.53 meV/V^2$ is the voltage-dependent polarizability as calculated in Ref. \cite{LiranPRX}.
The transmissive state requires a voltage amplitude modulation that induces an energy shift smaller by  $\Omega(V_G^{on})$, compared to the blocking voltage, $V_G^{\text{on}} < V_G^{\text{off}}$. 

Inverting the equations, the minimal transmission for polariton injected in energy $E_{LP0}(\beta)$ is given at
\begin{equation}
    V_{G}^{off} = \sqrt{\frac{E_{hh}-E_{LP0}(\beta)}{\alpha'}}.
\end{equation} 
Maximum transmission is expected at 
\begin{equation}
    V_{G}^{\text{on}}=\sqrt{\frac{E_{hh0}-(E_{LP0}(\beta)+\Omega(V_G^{on}))}{\alpha'}}.
\end{equation} 

The voltage modulation is the difference $V_G^{\text{on}} - V_G^{\text{off}}$ and is given by:
\begin{equation}
    \Delta V_{G}= V_G^{off}\left (1-\sqrt{\frac{E_{hh0}-(E_{LP0}(\beta) +\Omega(V_G^{on}))}{E_{hh0}-E_{LP0}(\beta)}} \right).
    \label{eq:SM:DV}
\end{equation}
We note that the Rabi frequency $\Omega(V_G)$ decreases with increasing voltage \cite{Rosenberg2016}, and can be described by $\Omega(V_G) = \Omega(V_G=0) \times \exp(-V_G^2 / 2\sigma_\Omega^2)$, where $\sigma_\Omega$ and $\Omega(V_G=0)$ are fitting parameter for our system, with values of 11~meV and 4~V, respectively.  
The experimental data and the model are shown together in Fig. \ref{fig:Fig1}(h).

\section{Energy per Operation Calculation}
\begin{figure}[h!]
    \centering
    \includegraphics[width=0.5\linewidth]{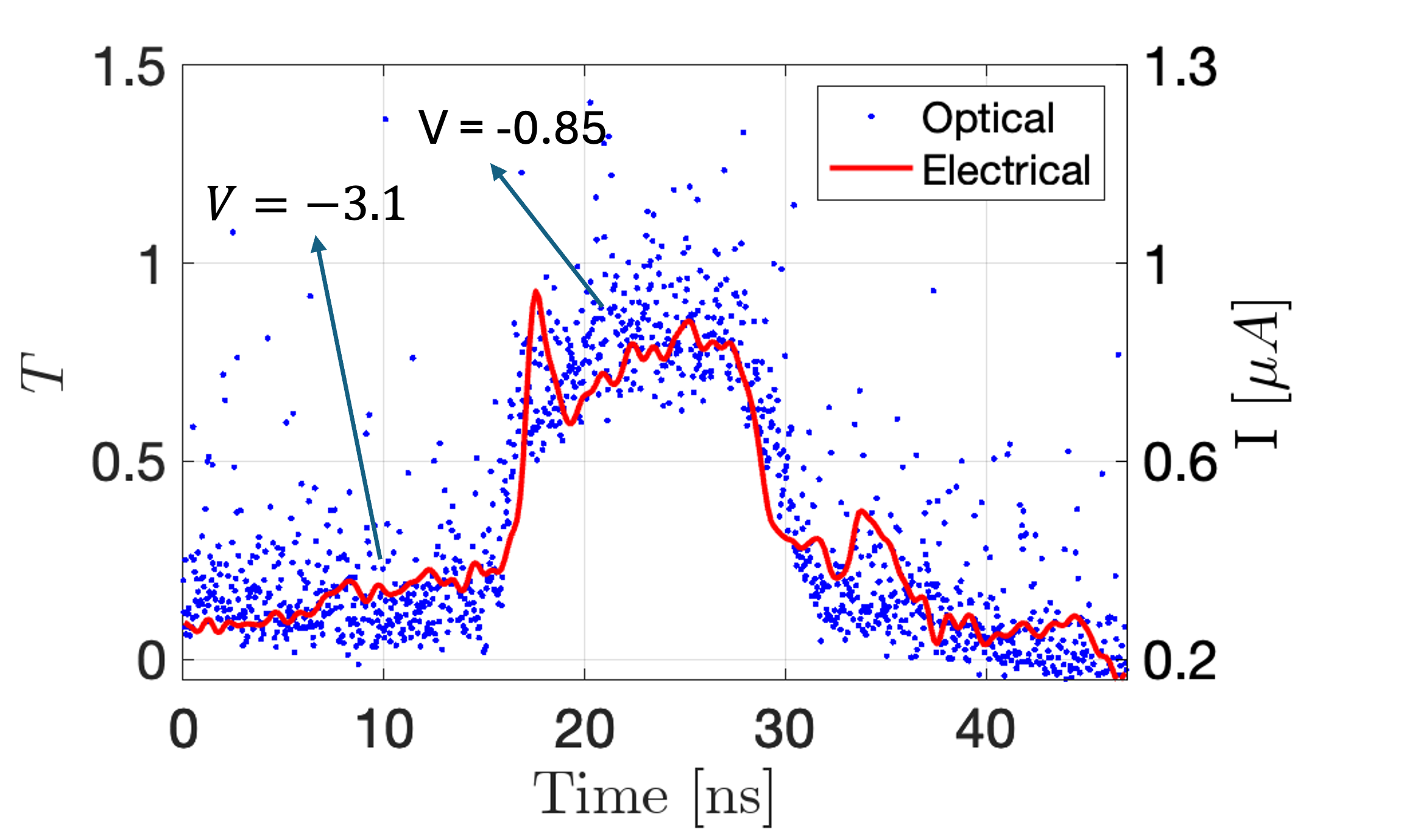}
    \caption{Time-resolved measurement of the electrical current at $E = -10.5 meV$  (\textit{red line}) and optical transmission (\textit{blue dots}) as a function of time. The off voltage is $V = -3.1 \, \text{V}$ with a ramp of $\Delta V = 2.25 \, \text{V}$ for transmission. This data was used to calculate the average energy per operation. The optical signal tracks the changes in the system, demonstrating the correlation between electrical and optical properties.}
    \label{SM:fig:Electrical_energy}
\end{figure}

To estimate the energy per operation, we consider two scenarios. (1) The device is in an \textit{on}-state all the time. This leads to a constant current (\textit{DC}) of $I_{on} = 0.46 \, \mu\text{A}$ at $V_{on} = -3.1 \, \text{V}$. (2) The device is modulated between the \textit{off} and \textit{on} states with a 50\%-50\% duty cycle $f = 1 \, \text{GHz}$. Here the current alternates between $I_{on},V_{on}$ and $I_{off} = 0.85 \, \mu\text{A}$ at $V_{off} = -0.85 \, \text{V}$. For the \textit{DC case}, the energy per operation is $U_{\text{DC}} = 1.43 \, \text{fJ}/\text{opreation}$ , at the 50\%-50\% duty cycle, the average current is $I_{\text{avg}} = \frac{I_{on} + I_{off}}{2}$, and the average voltage is $V_{\text{avg}} = \frac{V_{on} + V_{off}}{2}$. The energy per operation is then given by $E_{\text{op,avg}} = 1.29 \, \text{fJ}/\text{opreation}$. The data on which these calculations are based are plotted in Fig.- \ref{SM:fig:Electrical_energy}.

The nonlinear response of the system, as reported in Ref.~\cite{LiranPRX}, requires $\sim5000$ photons for a 500~ps pulse, a 0.5 duty cycle at a 1~GHz repetition rate, to achieve the same DWEP density required for the transistor operation. The energy per pulse, $E_{\text{pulse}}$, can be calculated as the product of the photon energy ($E_{\text{ph}}$) and the number of photons per pulse ($N_{\text{ph}}$):
\begin{equation}
    E_{\text{pulse}} = E_{\text{ph}}N_{\text{ph}}.
\end{equation}
For photons with an energy of approximately 1.527~eV, this corresponds to a total of 1.2~fJ per cycle, highlighting the system's remarkable energy efficiency.

\section{Optical Setup}

\begin{figure}[h!]
    \centering
    \includegraphics[width=0.8\linewidth]{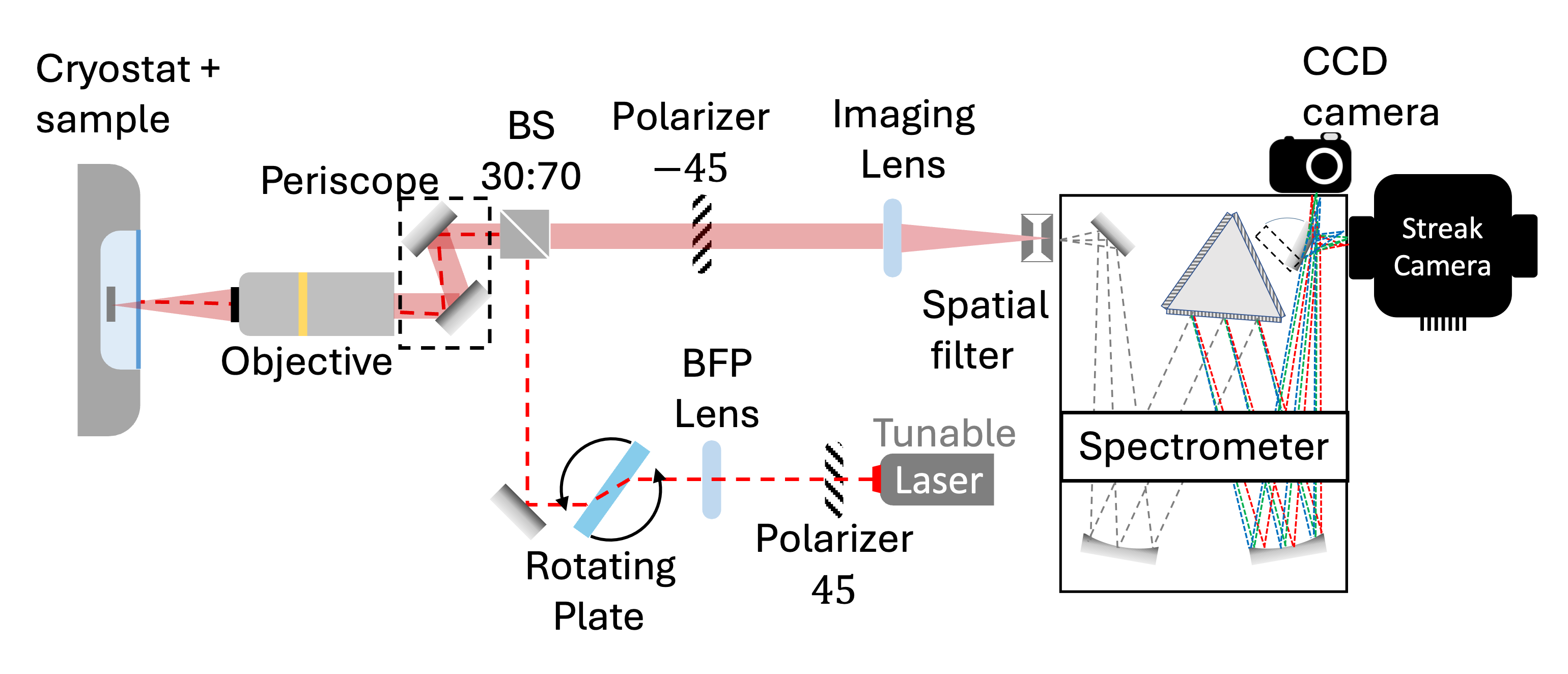}
    \caption{\textbf{The optical Setup:}
    A red dashed line marks the incident light, and a shaded red line marks the emitted light.}
    \label{sm:fig:setup}
\end{figure}

The optical setup, presented in Fig. \ref{sm:fig:setup}, contains a CW Titanium-Sapphire Laser  (model 3900 by spectra-Physics). The laser is focused on the back-focal plane (BFP) of the objective lens, at a spatial location such that it will enter the in-coupling grating at a desired single angle of incidence corresponding to the LP dispersion at a given energy. To choose the angle, we use a glass plate (about 5 mm thick), which deflects the beam to different positions at the BFP, i.e., different angles of incidence.
Furthermore, the laser light is polarized at a  \(45^{\circ}\) with respect to the in-coupling grating. The emission path includes cross-polarized light emitted at a  \(-45^{\circ}\) with respect to the out-coupling grating. This setup is designed to separate the emitted light from the scattered laser light.
After the light is emitted from the sample, it is imaged in the entrance slit of a spectrometer. A spatial filter is used to filter out all other emission except from the output grating coupler. The spectrometer has two outputs: a cooled CCD camera and a Streak Camera for imaging or time-resolved experiments respectively.

\section*{Data sets}
\begin{figure}[h]
    \centering
    \includegraphics[width=\linewidth]{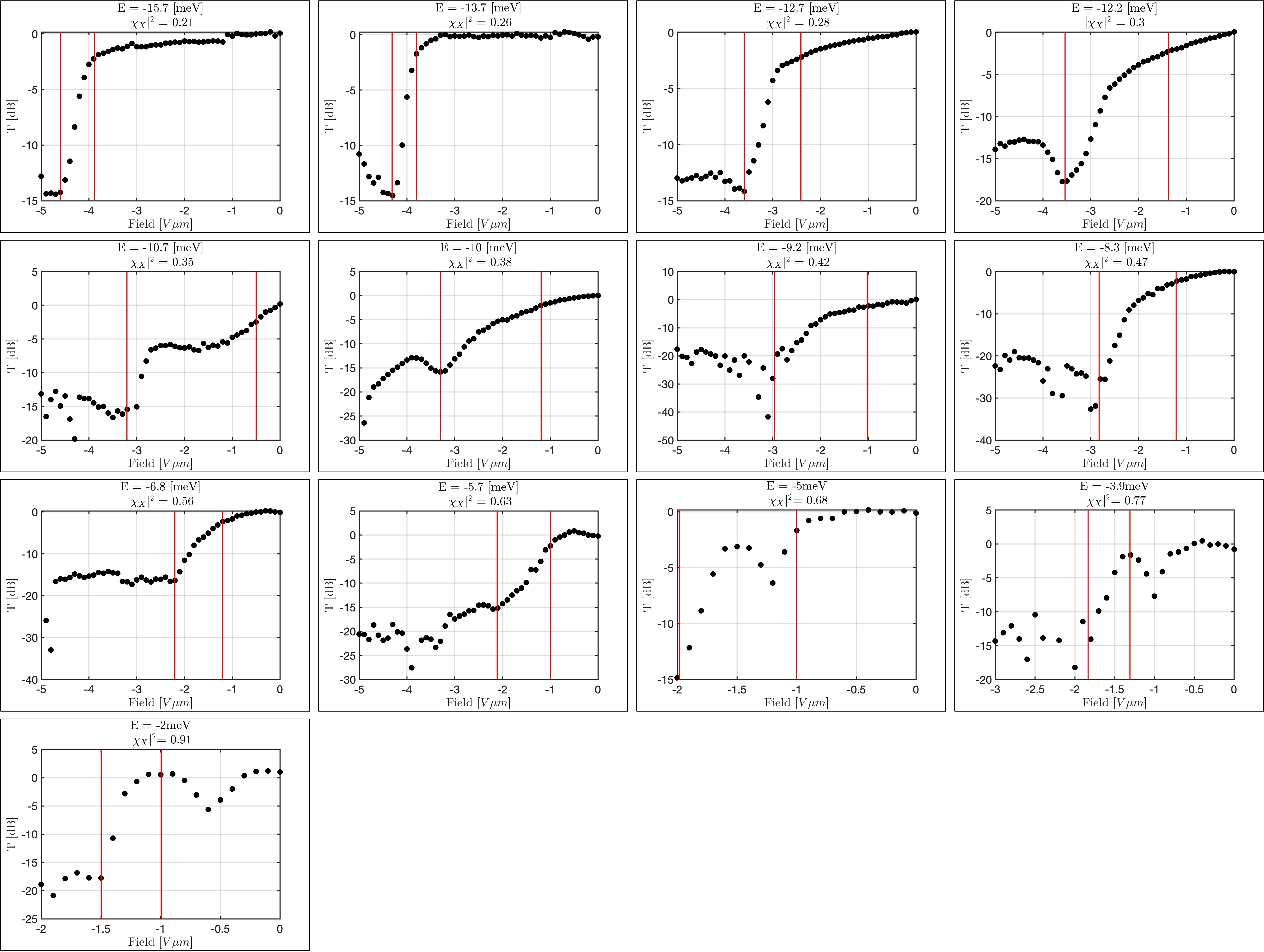}
    \caption{Raw Data of the switching performed with DC electric field, used in Fig. \ref{fig:Fig1}(g and h)}
    \label{fig:SM:DC_switch}
\end{figure}

\begin{figure}[h]
    \centering
    \includegraphics[width=0.75\linewidth]{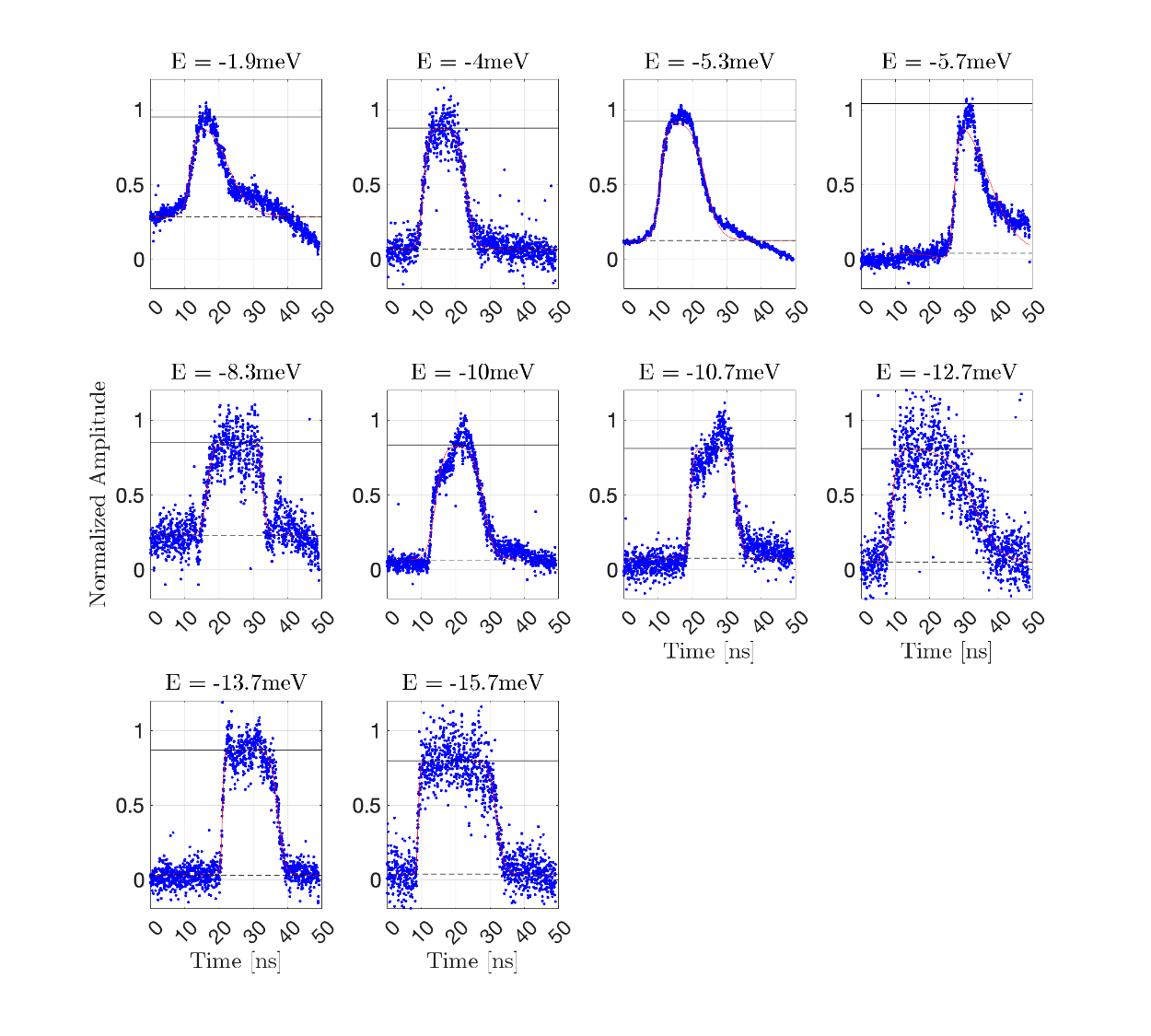}
    \caption{Raw Data of the switching performed with a pulsed electric field, used in Fig. \ref{fig:Fig3}(e-f)}
    \label{fig:SM:Pulses}
\end{figure}

Below, we plot the raw data used in the main text and data that supports the measurements in the main text.
Figures \ref{fig:SM:DC_switch} and \ref{fig:SM:Pulses} shows the Raw-data used for the analysis presented in Figures \ref{fig:Fig1} and \ref{fig:Fig3}.

